\documentclass[amsmath,amssymb, twocolumn,prl]{revtex4-1}
\usepackage{xcolor}
\usepackage{comment}
\usepackage{tikz, adjustbox}
\usetikzlibrary{arrows.meta, bending}
\newcommand{\nicesection}[1]{
\vspace{0em}
\begin{center}
\noindent\textbf{#1}
\end{center}
\vspace{0em}
}
\usepackage{microtype}

\usepackage{soul}

\begin{document}

\title{Unification of Decoupling Limits in String and M-theory}

\author{Chris D. A. Blair${}^1$, Johannes Lahnsteiner${}^2$, Niels A. Obers${}^{2, 3}$, and Ziqi Yan${}^2$ \smallskip\smallskip}

\affiliation{\smallskip%
${}^1$Instituto de F\'{i}sica Te\'{o}rica UAM/CSIC, Universidad Aut\'{o}noma de Madrid, Cantoblanco, Madrid 28049, Spain \smallskip\\
${}^2$Nordita, KTH Royal Institute of Technology and Stockholm University,
Hannes Alfv\'{e}ns v\"{a}g 12, SE-106 91 Stockholm, Sweden \smallskip\\
${}^3$Niels Bohr International Academy, The Niels Bohr Institute, Copenhagen University, Blegdamsvej 17,
DK-2100 Copenhagen \O, Denmark
}

\begin{abstract}

We study and extend the duality web unifying different decoupling limits of type II superstring theories and M-theory.
We systematically build connections to different corners, such as Matrix theories, nonrelativistic string and M-theory, tensionless (and ambitwistor) string theory, Carrollian string theory, and Spin Matrix limits of AdS/CFT. 
We discuss target space, worldsheet, and worldvolume aspects of these limits in arbitrary curved backgrounds.

\end{abstract}

\maketitle

\begin{figure*}
\centering
\begin{adjustbox}{width=.95\textwidth}
\hspace{-8mm}
\begin{tikzpicture}
    \begin{scope}[scale=.8]

        \path[every node/.style=draw, rounded corners=1, line width=1.5pt, minimum width=135pt, minimum height=20pt, font = \small]   
                (3,1.5) node[fill=gray!20] {DLCQ M-Theory}
                (11,1.5) node {\textbf{M0T}: BFSS Matrix theory}
                (19,1.07) [align=center] node[minimum height = 40pt] {\hspace{4mm} IKKT Matrix theory\\
                \hspace{4mm} tensionless string theory\\
                \hspace{4mm} ambitwistor string theory}
                (3,3.5) node {IIB nonrelativistic string theory}
                (11,3.5) node {\textbf{M1T}: Matrix string theory}
                (19,3.5) node {\textbf{MM0T}: Spin Matrix theory}
                (3,5.45) node[fill=gray!20] {nonrelativistic M-Theory}
                (11,5.5) node {\textbf{M}atrix \textbf{p}-brane \textbf{T}heory}
                (11,0) node {Carrollian string theories} 
                (19,5.5) node {\textbf{M}ulticritical \textbf{M(p\,-1)T}};

        \path   (7,1.8) [font=\footnotesize] node {\emph{lightlike}}
                (7,1.25) [font=\footnotesize] node {\scalebox{0.7}{\emph{compactification}}}
                (7,3.25) [font=\footnotesize] node {\emph{S-dual}}
                (4.8,4.5) [font=\footnotesize] node {\,\scalebox{0.85}{\emph{toroidal compactification}}}
                %
                (15,1.8) [font=\footnotesize] node {\emph{timelike}}
                (15,1.25) [font=\footnotesize] node {\emph{T-dual}}
                (15,3.2) [font=\footnotesize] node {\emph{DLCQ}}
                (15,5.2) [font=\footnotesize] node {\emph{DLCQ}}
                (11.65,4.5) [font=\footnotesize] node {\emph{T-dual}}
                (11.65,2.47) [font=\footnotesize] node {\emph{T-dual}}
                (19.65,4.5) [font=\footnotesize] node {\emph{T-dual}}
                (19.65,2.47) [font=\footnotesize] node {\emph{DLCQ}}
                (-.7,3.5) [font=\footnotesize] node[rotate=90] {\emph{U-dual}}
                (16.6,1.09)[font=\small] node[rotate=90] {\textbf{M(-1\hspace{-0.5mm})T}}
                (15.1,0.05) [font=\footnotesize] node[rotate=27] {\emph{T-dual}};

        \draw[-{>[length=1.3mm]}] (6,1.5) -- (8,1.5);
        \draw[{<[length=1.3mm]}-{>[length=1.3mm]}] (6,3.5) -- (8,3.5);
        \draw[-{>[length=1.3mm]}, dashed] (6,5.5) -- (8,5.5);
        \draw [{<[length=1.3mm]}-{>[length=1.3mm]}] (14,1.5) -- (16,1.5);
        \draw [-{>[length=1.3mm]}] (14,3.5) -- (16,3.5);
        \draw [-{>[length=1.3mm]}] (14,5.5) -- (16,5.5);

        \draw [-{>[length=1.3mm]}] (3,5) -- (3,4);
        \draw [{<[length=1.3mm]}-{>[length=1.3mm]}] (11,4) -- (11,5);
        \draw [{<[length=1.3mm]}-{>[length=1.3mm]}] (11,2) -- (11,3);
        \draw [-{>[length=1.3mm]}] (19,2) -- (19,3);
        \draw [{<[length=1.3mm]}-{>[length=1.3mm]}] (19,4) -- (19,5);

        \draw [-{>[length=1.3mm]}] (6-0.05,5.03) -- (8,4);
        \draw [{<[length=1.3mm, quick]}-{>[length=1.3mm, quick]}] (14,-.5+0.5) .. controls (15,-.5+0.5) and (15,0+0.5) .. (16,0+0.6);
        
        \draw [{<[length=1.3mm]}-{>[length=1.3mm]}] (0,2) to[out=120,in=240] (0,5);

\end{scope}
\end{tikzpicture}
\end{adjustbox}
\caption{The duality web of decoupling limits. Here, DLCQ is short for the procedure of compactifying the theory over a lightlike circle followed by a T-duality. In the web, these lightlike circles become spacelike in their T-dual theories.}
\label{fig:rm}
\end{figure*}
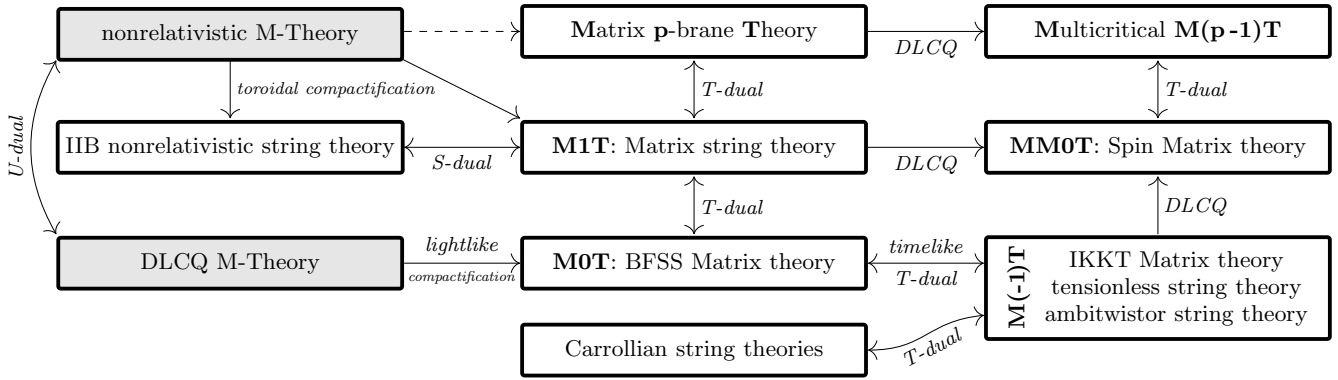

\nicesection{Introduction}

Faced with complicated physical theories, we often seek simplifying limits that still capture key behavior or structure. This logic was reversed in the introduction of M-theory, which was conjecturally defined by viewing `dual' string theory scenarios as limits of an underlying theory of quantum gravity \cite{Hull:1994ys,Witten:1995ex}. 
These limits include, as a starter, the perturbative superstring theories and the 10D and 11D supergravities.
Armed with this picture, we can search for other accessible limits of string/M-theory.

In this Letter, we consider a class of decoupling limits, which offer significant simplifications by removing many states from the spectrum, and sometimes allow us to have a peek into various nonperturbative regimes.
These limits can be viewed as generalizations of the nonrelativistic point particle limit.
To motivate this, consider the action for a charged point particle in $d$ dimensions:
\begin{equation}
    S = - m \int d \tau \, \sqrt{-\eta_{\mu\nu} \, \dot{x}^\mu \, \dot{x}^\nu } 
+ q \int d \tau \, A_{\mu} \, \dot{x}^\mu\,,
\label{relpp}
\end{equation}
where $\mu = 0\,, \, \cdots, \, d-1$.
To take the nonrelativistic limit, we (re)introduce the factor of $c$ such that $m \rightarrow m \, c$ and $\eta_{\mu\nu} \rightarrow (-c^2 , \delta_{ij})$ with $i\,, j = 1, \cdots, \, d-1$. When we let $c \rightarrow \infty$ the action is naively divergent due to the term involving the rest mass. 
For the special class of \emph{BPS particles} whose mass equals their charge, $m=q$, we can cancel this divergence in a universal way by introducing a \emph{critical} gauge field of the form $A = c \, d x^0$.
Then, expanding the action \eqref{relpp} gives the nonrelativistic particle action:
\begin{equation}
S = \tfrac{m}{2} \int d \tau \,\dot{x}^i \, \dot{x}^i
\,,
\label{nonrelpp}
\end{equation}
where we let $x^0=\tau$.
An important class of particles whose mass equals their charge are those arising as Kaluza-Klein excitations on reduction from an extra periodic dimension. 
In this case, the nonrelativistic limit can be ``baked in'' to the form of the higher-dimensional geometry, by taking the metric of the latter to have the form
\begin{equation}
d s^2 = \tfrac{R^2}{c^2} \Bigl( d y + \tfrac{c^2}{R} d x^0 \Bigr)^2 - c^2 (dx^0)^2
+ dx^i d x^i \,.
\label{nonrelppUplift}
\end{equation}
where $y \sim y+ 2\pi$. 
Dimensional reduction of the massless particle in $d+1$ dimensions gives the action \eqref{relpp} with $m=q=\tfrac{Nc}{R}$\,, with $N$ the Kaluza-Klein (KK) number and the KK vector providing the critical gauge field $A= c \, d x^0$. 
In the $(d+1)$-dimensional geometry of equation \eqref{nonrelppUplift}, the limit $c \rightarrow \infty$ means that the extra circular dimension becomes lightlike.
This defines the Discrete Light Cone Quantization (DLCQ) of the higher-dimensional theory. 

M-theory in the DLCQ can be defined using a closely related limit, which is usually interpreted as an infinite boost from a spatial compactification~\cite{Seiberg:1997ad, Sen:1997we}. In this limit, all the light excitations except the bound states of $N$ nonrelativistic D0-particles decouple, and their dynamics is described by Banks-Fischler-Shenker-Susskind (BFSS) Matrix theory that generalises the simple action~\eqref{nonrelpp}~\cite{deWit:1988wri, Banks:1996vh}. In the large $N$ limit, BFSS is conjectured to describe the whole M-theory in flat spacetime.  
The DLCQ of quantum field theory can be problematic~\cite{Hellerman:1997yu,Chapman:2020vtn}. However, the DLCQ of string or M-theory is believed to be well defined since it can be given a first-principles definition using T- (or U-)duality, where the dual side is a self-consistent theory with a conventional spatial compactification~\cite{Gomis:2000bd, Danielsson:2000gi, Bergshoeff:2018yvt, Ebert:2023hba}. 

This Letter aims to systematically explore generalizations of the above classic example, and its DLCQ M-theory version.
These generalizations involve nonrelativistic decoupling limits associated not simply with BPS particles but with strings and branes. 
We exploit string/M-theory dualities to map different such decoupling limits into each other, and further combine these dualities with additional DLCQ limits.
In doing so, we consider the curved background analogs of the nonrelativistic limits. For instance, in the example above, this amounts to covariantizing the flat spacetime limit via $dx^0 \rightarrow \tau_\mu \, d x^\mu$, $d x^i \rightarrow E_\mu{}^i \, dx^\mu$, where $\tau$ and $E^i$ are orthogonal {Vielbeine} in Newton-Cartan geometry \emph{i.e.} the covariant formalism of Newtonian gravity \cite{Hartong:2022lsy}.

Our results update and extend the known features of the duality web surrounding DLCQ M-theory~\cite{Dijkgraaf:1997vv, Gopakumar:2000ep, Harmark:2000ff, Gomis:2000bd, Danielsson:2000gi, Hyun:2000cw}, taking into account recent developments in curved non-Lorentzian geometries \cite{Harmark:2017rpg,Bergshoeff:2018yvt,Oling:2022fft}, which appear as the background geometry after taking the limit.
We will see that this duality web
encompasses not only all Matrix theories, but also different string theories that are 
\emph{nonrelativistic}~\cite{Gomis:2000bd, Danielsson:2000gi}, \emph{tensionless}~\cite{Lindstrom:1990qb}, \emph{ambitwistor}~\cite{Mason:2013sva}, and \emph{Carrollian}~\cite{Cardona:2016ytk}. We will also reveal novel corners that arise from a \emph{2nd} DLCQ, which are relevant for \emph{Spin Matrix theories} (SMT) from near-BPS limits of AdS/CFT \cite{Harmark:2017rpg}. The 11D uplifts of all these limits give rise to different corners of M-theory that are U-dual to each other. See Fig.~\ref{fig:rm} for a road map of this duality web.

\nicesection{Matrix Theory from Critical Fields}


{Having discussed the nonrelativistic particle limit, we now proceed to describe its stringy generalization that produces \emph{nonrelativistic string theory} (NRST) \cite{Klebanov:2000pp, Gomis:2000bd, Danielsson:2000gi} (see \cite{Oling:2022fft} for a review), a unitary and ultraviolet complete string theory. 
In 10D spacetime with coordinates $x^\mu$, $\mu = 0\,,\cdots,9$\,, we focus on the bosonic sector and write the background (string frame) metric $G_{\mu\nu}$\,, $B$-field, dilaton $\Phi$ and Ramond-Ramond (RR) forms $C_q$~\cite{Ebert:2021mfu} as
\begin{subequations} \label{eq:nrsl}
\begin{align}
    B & = - \omega^2 \, \tau^0 \wedge \tau^1 + b\,, 
        \hspace{0.68cm}
    \tau^A = \tau_\mu{}^A \, dx^\mu, \\[2pt]
    G^{}_{\mu\nu} & = \omega^2 \, \tau^{}_{\mu\nu} + E^{}_{\mu\nu}\,, 
        \hspace{1.25cm}%
    \Phi = \varphi + \ln \omega\,,\\[4pt]
    C_q & = \omega^2 \, \tau^0 \wedge \tau^1 \wedge c_{q-2} + c_q \,.
\end{align}
\end{subequations}
Here, the two-form $B$-field becomes critical, and $\omega$ corresponds to the ratio between the speed of light and string velocity \cite{Hartong:2021ekg}.
We will focus on type IIB with even $q$.
We define NRST as $\omega \rightarrow \infty$: 
just as in the particle case, the $\omega^2$ divergences cancel in the F1-string action due to the BPS nature of the string.
NRST has a Galilean invariant string spectrum in flat background, hence the name ``nonrelativistic" \cite{Gomis:2000bd}.  The modern understanding \cite{Andringa:2012uz, Harmark:2017rpg, Bergshoeff:2018yvt,Bergshoeff:2021bmc,Bidussi:2021ujm,Ebert:2021mfu} is that the target space geometry is non-Lorentzian, described by the longitudinal and transverse Vielbeine $\tau_\mu{}^A$, $A = 0\,, 1$ and $E_\mu{}^{A'}$, $A' \! = 2\,, \cdots, 9$\,, respectively. In \eqref{eq:nrsl} we defined
\begin{equation} \label{eq:defte}
    \tau^{}_{\mu\nu} = \tau^{}_\mu{}^A\, \tau^{}_\nu{}^B \, \eta^{}_{AB}\,,
        \qquad%
    E^{}_{\mu\nu} = E^{}_\mu{}^{A'}  E^{}_\nu{}^{B'} \, \delta^{}_{A'B'}\,.
\end{equation}
These Vielbeine are related by \emph{string Galilei boosts} 
$\delta^{}_\text{G} \tau^A = 0$ and $\delta^{}_\text{G} E^{A'} = \lambda^{A'}{}_{\!\!A} \, \tau^A$
that naturally generalizes the Galilean boosts in Newton-Cartan geometry. NRST also couples to the background $B$-field $b$\,, dilaton $\varphi$ and RR forms $c_q$\,. The Galilean boosts can also act nontrivially on background gauge potentials~\cite{Bergshoeff:2021bmc, Bidussi:2021ujm}. 
We emphasize that when we write a background such as \eqref{eq:nrsl} and \eqref{eq:defte}, which should be viewed as the prescription for taking a particular decoupling limit, the surviving fields (here $\tau^A$, $E^{A'}$\!, $b$\,, $c_q$ and $\varphi$) are those that appear in the theory that follows from the $\omega \rightarrow \infty$ limit.

S-duality\,\footnote{See~\cite{Bergshoeff:2022iss,Bergshoeff:2023ogz} for its SL($2\,, \mathbb{Z}$) generalization in IIB NRST, which exhibits interesting structure.} maps IIB NRST to a different decoupling limit of relativistic IIB, defined by the infinite $\omega$ limit of the reparametrization~\cite{Ebert:2023hba, longpaper}, 
%
\begin{subequations} \label{eq:1bl}
\begin{align}
    C_2 & = \omega^2 \, e^{-\varphi} \, \tau^0 \wedge \tau^1 + c_2\,, \\[2pt]
    G^{}_{\mu\nu} & = \omega \, \tau^{}_{\mu\nu} + \omega^{-1} E^{}_{\mu\nu}\,,  
        &%
    \Phi &= \varphi - \ln \omega\,, \label{eq:1blgp} 
\end{align}
\end{subequations}
derived by S-dualizing \eqref{eq:nrsl}. The background fields $B = b$ and $C_q = c_q$ with $q \neq 2$ are unchanged in this limit. Now, the RR 2-form $C_2$ instead of the $B$-field is critical. We name the theory obtained in the $\omega \rightarrow \infty$ limit \emph{Matrix 1-brane Theory} (M1T), which we justify around \eqref{eq:dpdl}. 

T-dualizing M1T in spatial directions gives a tower of critical RR ($p$\,+1)-form limits, where \eqref{eq:1bl} becomes
\begin{subequations} \label{eq:pbl}
\begin{align}
    C_{p+1} & = \omega^2 \, e^{-\varphi} \, \tau^0 \wedge \cdots \wedge \tau^{p} + c_{p+1}\,, \qquad\,\, p\geq 0\,, \\[2pt]
    G^{}_{\mu\nu} & = \omega \, \tau^{}_{\mu\nu} + \omega^{-1} E^{}_{\mu\nu}\,, 
        \qquad%
    \Phi = \varphi + \tfrac{p-3}{2} \ln \omega\,.  \label{MpT}
\end{align}
\end{subequations}
In $\omega\rightarrow\infty$, we have \emph{Matrix $p$-brane Theory} (M$p$T), where the spacetime geometry has a codimension-$(p+1)$ foliation, described by the longitudinal $\tau^A$ with $A = 0\,, \cdots, p$ and transverse $E^{A'}$ with $A' \! = p+1\,, \cdots, 9$ that are related via a $p$-brane Galilei boost. These Vielbein fields define the non-Lorentzian \emph{$p$-brane Newton-Cartan geometry}. Here, T-duality maps between longitudinal and transverse directions as well as the critical RR fields. 
Such limits and their dualities date back to the original studies of e.g.~\cite{Gopakumar:2000ep,Harmark:2000ff, Hyun:2000cw} focusing on 
``open string'' decoupling limits in the presence of particular branes and
\cite{Gomis:2000bd, Danielsson:2000gi} highlighting the ``closed string'' limits independent of the presence of branes.
In contrast, the $\omega\rightarrow\infty$ limit here is applied to the full type II string theory containing all possible extended objects in general backgrounds. 

The light excitations in the critical RR ($p$\,+1)-form limit are $N$ D$p$-brane bound states. 
Taking the limit at the level of the non-Abelian D$p$ worldvolume action \cite{Myers:1999ps} (for simplicity we omit fermions and take the target-space $q$-form gauge fields to vanish) we get:
\begin{align} \label{eq:dpdl}
    & S^{}_{\text{D}p} = - \tfrac{T^{}_p}{2} \! \int \! d^{p+1} \sigma \, e^{-\varphi} \, \text{Tr} \biggl[ \sqrt{-\tau} \, \Bigl( \tau^{\alpha\beta} \, D_\alpha \Phi^{\mu} \, D_\beta \Phi^{\nu} \, E_{\mu\nu} \notag \\[2pt]
    & + \tfrac{1}{2} \tau^{\alpha\gamma} \tau^{\beta\delta} F^{}_{\alpha\beta} F^{}_{\gamma\delta} - \tfrac{1}{2} \, \bigl[\Phi^{\mu}\!, \Phi^{\nu}\bigr] \bigl[\Phi^{\rho}\!, \Phi^{\sigma}  \bigr] E_{\mu\rho} \, E_{\nu\sigma} \Bigr) \biggr],\!
\end{align}
where $\tau_{\alpha\beta}$ is the pullback of $\tau_{\mu\nu}$ in \eqref{MpT} to the worldvolume
(in static gauge) with $\tau^{\alpha\beta}$ the inverse and $\tau = \det \tau_{\alpha\beta}$. The adjoint scalar $\Phi^{\mu}$ describes 
the fluctuation of the D$p$-branes in the transverse directions, $F_{\alpha\beta}$ is the SU($N$) field strength and $D_\alpha$ is the covariant derivative. The open string gauge potential is never changed by the $\omega$ reparametrization. 
In flat spacetime, the D$p$-brane action \eqref{eq:dpdl} describes Matrix theory compactified over a $p$-torus, giving BFSS Matrix theory in M0T, Matrix string theory \cite{Dijkgraaf:1997vv} in M1T and general Matrix gauge theories in M$p$T. The name M$p$T is precisely justified as its light excitations are D$p$-branes described by various Matrix theories.  
The $p=0$ case corresponds to the example of the Introduction.


\nicesection{Multicriticality and Spin Matrix Theory}

T-dualizing a longitudinal spatial circle in type IIB NRST gives rise to the DLCQ of type IIA string theory, where the latter contains a lightlike circle that T-dualizes to a well-defined spacelike circle in IIB NRST. This allows one to use the self-consistent NRST as a first-principles definition of DLCQ string theory, avoiding having to deal directly with the lightlike circle.
All the limits we have considered so far have therefore been connected by (spacelike) T-dualities and S-dualities to the DLCQ of IIA.
We can now extend the duality web by taking a DLCQ of any of these limits and further dualizing. 

T-dualizing the lightlike circle in DLCQ IIB NRST gives back DLCQ IIA NRST, where the dual circle is still lightlike~\cite{Bergshoeff:2018yvt, Ebert:2021mfu}. 
The situation changes, and becomes much richer, with S-duality in play which maps DLCQ IIB NRST to DLCQ M$1$T.
In this case, the lightlike circle in DLCQ M$1$T maps under T-duality to a spacelike circle. 
Hence, fascinatingly, DLCQ M$1$T receives a first-principles definition from its T-dual theory.
T-duality in a lightlike isometry of \eqref{eq:1bl} for M$1$T gives
\begin{subequations} \label{eq:smtrp}
\begin{align}
    B & = - \omega \, \tau^0 \wedge \tau^1 + b\,, 
        \qquad%
    C_1 = \omega^2 \, e^{-\varphi} \, \tau^0 + c_1\,, \\[2pt]
    \Phi & = \varphi - \ln \omega\,,
        \qquad%
    C_q = \omega \, \tau^0 \wedge \tau^1 \wedge c_{q-2} + c_q\,, \\[2pt]
    G^{}_{\mu\nu} & = - \omega^2 \, \tau^{}_\mu{}^0 \, \tau^{}_\nu{}^0 + \tau^{}_\mu{}^1 \, \tau^{}_\nu{}^1 + \omega^{-1} \, E^{}_{\mu\nu}\,. \label{eq:smtg}
\end{align}
\end{subequations}
In the $\omega \rightarrow \infty$ limit, both $B$ and $C_1$ are critical. The lightlike circle in DLCQ M1T maps to the spatial $x^1$ isometry, associated with the Vielbein $\tau^1$ of weight $\omega^0$. We refer to this new theory defined by the $\omega \rightarrow \infty$ limit of \eqref{eq:smtrp} as \emph{Multicritical Matrix 0-brane Theory} (MM0T).
Similar to the T-dual relations between M$p$Ts, T-dualizing transverse isometries in MM0T gives \emph{Multicritical Matrix $p$-brane Theory} (MM$p$T) \cite{longpaper, Gomis:2023eav} where
%
the light excitations are F1-D$p$ bound states.
Heuristically, this limit corresponds to a critical F1-D$p$ background. 
T-dualizing a longitudinal $x^1$ isometry asscociated with the F1 direction 
in MM$p$T leads to DLCQ M$p$T. 
MM$p$T can be viewed as the DLCQ of the DLCQ of IIA/IIB, while M$p$T is the DLCQ of IIA/IIB. 

%
The Polyakov action of the MM$0$T F1-string is~\cite{Gomis:2023eav}
\begin{align} 
    S_{\text{MM$0$T}} = - \tfrac{T}{2} \! \int \! d^2 \sigma \, \Big( & - \lambda_A \, \partial_\sigma X^\mu \, \tau_\mu{}^A + \lambda_1 \, \partial_\tau X^\mu \, \tau_\mu{}^0 \notag \\[2pt]
    & + \partial_\sigma X^\mu \, \partial_\sigma X^\nu E_{\mu\nu}
    \Big) - T \! \int \! b\,.
\label{SMTws} 
\end{align}
Here, 
$\lambda_A$ with $A = 0\,, 1$ are Lagrange multipliers. This action is written in the conformal gauge with the worldsheet coordinates $(\tau, \sigma)$\,. The MM$0$T string is invariant under the worldsheet Galilei boost $\delta_\text{G} \tau = 0$\,, $\delta_\text{G} \sigma = v \, \tau$ and proper transformations of $\lambda_A$. In this sense, MM$0$T has a nonrelativistic worldsheet. This is due to the backreaction of uncancelled $\omega$ divergences, which we have rewritten in~\eqref{SMTws} using Hubbard-Stratonovich transformations by introducing Lagrange multipliers as in \cite{Gomis:2000bd}.

The worldsheet action \eqref{SMTws} generalizes the nonrelativistic string
that appears in the \emph{Spin Matrix Theory} (SMT) limits of AdS/CFT~\cite{Harmark:2017rpg}. 
SMTs are integrable quantum mechanical theories that arise from the near BPS limits of $\mathcal{N}=4$ SYM in the regime of almost zero 't Hooft coupling. The SU($1,2|3$) SMT limit corresponds to the M$1$T-type limit of the bulk AdS${}_5 \times S^5$ metric with a lightlike isometry. T-dualizing this lightlike isometry leads to the $\omega \rightarrow \infty$ limit of the reparametrized metric~\eqref{eq:smtg} in MM$0$T. 
In particular, the MM0T string action \eqref{SMTws} matches the SMT string action (A.15) in \cite{Harmark:2018cdl} (see also~\cite{Bidussi:2023rfs} for the latest developments on the SMT string).

\nicesection{D-Instanton Limit and Tensionless String} \label{sec:dil}

We have seen that BFSS Matrix theory lives on the D0-particles in IIA M$0$T. 
The IIB \emph{Ishibashi-Kawai-Kitazawa-Tsuchiya (IKKT) Matrix theory} \cite{Ishibashi:1996xs} arises from the timelike T-duality of BFSS. Therefore, IKKT describes the light D($-$1)-brane (\emph{i.e.}, D-instanton) dynamics in M(-1)T, which arises from a critical RR 0-form limit of IIB${}^*$ string theory (the timelike T-dual of IIA \cite{Hull:1998vg}). The IIB$^*$ string theory is identical to the IIB theory except the dilaton $\Phi$ in IIB$^*$ gains a shift $i \, \pi / 2$\,, resulting in an imaginary string coupling. This shift in the dilaton can be exchanged with making the RR charges imaginary.
  
T-dual of M0T in a timelike isometry defines M(-1)T, 
\begin{align} \label{eq:nobl}
    C_{0} & = \tfrac{\omega^2}{e^{\varphi}} + c_{0},
        &%
    G^{}_{\mu\nu} & = \tfrac{\tau^{}_{\mu\nu}}{\omega}, 
        &%
    \Phi & = \varphi + 
    \tfrac{i\pi}{2} -2 \, \ln \omega. 
        %
\end{align}
Here $\tau_{\mu\nu}$ is defined as in \eqref{eq:defte} but with $A = 0\,, \cdots, 9$ \footnote{Here (and below) we swap the notation $\tau$ and $E$ such that the former always has Lorentzian signature and the latter Euclidean.
}. It is understood that other background fields are not rescaled. Applying \eqref{eq:nobl} to the action of a stack of D-instantons in IIB reproduces IKKT for $\omega \rightarrow \infty$.  

The F1-string action in M(-1)T describes a tensionless string~\cite{Lindstrom:1990qb, Isberg:1993av}. Plugging $G_{\mu\nu} = \omega^{-1} \, \tau_{\mu\nu}$ into the (bosonic) Nambu-Goto string action with $B=0$, we get
\begin{equation} \label{eq:sfs}
    S_\text{F1} = - \tfrac{T}{\omega} \int d^2 \sigma \, \sqrt{- \det \bigl( \partial_\alpha X^\mu_{\phantom{\dagger}} \, \partial_\beta X^\nu_{\phantom{\dagger}} \, \tau_{\mu\nu} \bigr)}\,. 
\end{equation}
Its $\omega \rightarrow \infty$ limit is equivalent to the tensionless~\cite{Lindstrom:1990qb} (or Gross-Mende~\cite{Gross:1987ar}) limit. In \cite{Lindstrom:1990qb}, it is shown that this tensionless limit results in a finite action with a nonrelativistic worldsheet~\cite{Bagchi:2013bga}. 
Note that such nonrelativistic string worldsheets also show up in M$p$Ts, which are T-dual to M(-1)T, as well as in MM$p$T~\cite{Harmark:2018cdl, Bidussi:2023rfs, Gomis:2023eav}.

We have seen that the D-instantons and F1-strings in M(-1)T are associated with IKKT and tensionless strings, respectively. 
It is also interesting to note that there are three known vacua in tensionless string theory that lead to distinct quantum theories \cite{Siegel:2015axg, Casali:2016atr, Bagchi:2020fpr}. One such vacuum gives the ambitwistor string, which is related to the Cachazo-He-Yuan (CHY) formula of field-theory amplitudes \cite{Cachazo:2013hca}. 
It would be interesting to revisit these relations in the broader context of M(-1)T, whose dynamics is largely encoded by IKKT on the D-instantons. 

\nicesection{Carrollian String Theory}

The (-1)-brane limit is a special case in a hierarchy of new decoupling limits of type II string theories that exhibit spacetime Carrollian behaviors. These decoupling limits are generated by T-dualizing $q$ spacelike directions in M(-1)T, which maps \eqref{eq:nobl} to
\begin{subequations} \label{eq:noblp}
\begin{align}
    C_{q} & = \omega^2 \, e^{-\varphi} \, E^1 \wedge \cdots \wedge E^{q} + c_{q}, \\[2pt]
    G^{}_{\mu\nu} & = \omega \, E^{}_{\mu\nu} + \omega^{-1} \tau^{}_{\mu\nu}\,, 
        \quad%
    \Phi \! = \! \varphi + \! \tfrac{i \, \pi}{2} \! + \! \tfrac{q-4}{2} \ln \omega.
        %
        %
\end{align}
\end{subequations}
Here, $\tau_{\mu\nu}$ and $E_{\mu\nu}$ are given as in \eqref{eq:defte} but with $A = 0\,, \, q+1\,, \cdots, \, 9$ and $A' = 1\,, \cdots, \, q$\,. 
The prescription~\eqref{eq:noblp} defines M$p$T for $p = -q -1 < -1$~\footnote{The M$p$Ts contain the Carrollian strings constructed in~\cite{Cardona:2016ytk}. See more in~\cite{Gomis:2023eav}.}. 
Relative to M$p$T for $p\geq0$\,, we have swapped $\tau$ and $E$ to keep the timelike index in $\tau$.
Note that $\tau_{\mu\nu}$ is of rank $10-q$ and $E_{\mu\nu}$ is of rank $q$\,. In $\omega \rightarrow \infty$, the metric description is invalid and $\tau^A$ and $E^{A'}$ are related via a \emph{$(9-q)$-brane Carrollian boost} $\delta^{}_\text{C} \tau^A = \lambda^{A}{}^{}_{\!A'} \, E^{A'}$ and $\delta^{}_\text{C} E^{A'} = 0$. 
This is the Carrollian boost for particles when $q = 9$\,: in flat spacetime, it acts on time $t$ and space $x^{A'}$ as $\delta_\text{C} t = v^{}_{\!A'} \, x^{A'}$ and $\delta_\text{C} x^{A'} = 0$~\cite{levy1965nouvelle, sen1966analogue, Duval:2014uoa}; the curved 
Carrollian geometry arises from a zero speed-of-light limit of relativity
~\cite{Hartong:2015xda,Hansen:2021fxi}. In general, the $\omega \rightarrow \infty$ limit of \eqref{eq:noblp} describes 
strings in generalized Carrollian geometry~\cite{Bergshoeff:2020xhv}. This framework provides a new perspective on Carrollian theories in connection to Matrix theories on branes localized in time that are T-dual to IKKT. It is tempting to speculate that this may shed light on Carrollian/celestial holography \cite{Donnay:2022aba}.

Timelike T-duality maps M$p$T to M(-$p$\,-1)T for all $p$ and MM$p$T to DLCQ M(-$p$\,-1)T for $p\geq0$\,. The light excitations of M$p$T for $p<-1$ are type II${}^*$ Euclidean branes (see also~\cite{Gutperle:2002ai} for spacelike branes), which can be viewed as tachyons \cite{Hull:1998vg}. This is natural in view of the fact that tachyons give rise to zero energy particles in Carrollian theories \cite{deBoer:2021jej}. 

In M(-3)T, the RR 2-form becomes critical. S-dualizing M(-3)T leads to the Carrollian analog of NRST where the $B$-field becomes critical and it cancels the background instantonic F1-string tension. The defining prescriptions for this \emph{Carrollian string theory} are
\begin{subequations}
\begin{align}
    B & = - i \, \omega^2 \, E^8 \wedge E^9 + b\,, \\[2pt]
    G_{\mu\nu} & = \tau_{\mu\nu} + \omega^2 \, E_{\mu\nu}\,,
        \qquad%
    \Phi = \varphi - \tfrac{i\pi}{2} + \ln \omega\,.
\end{align}
\end{subequations}
Here, $A = 0\,, \cdots, 7$ and $A' = 8\,, 9$\,. The $\omega \rightarrow \infty$ limit leads to the analog of F1-string action in NRST \cite{Oling:2022fft}, but now in Carroll-like target space geometry.

\nicesection{M-theory Uplifts}

Finally, we can extend the duality web to incorporate the decoupling limits of M-theory uplifting the string theories discussed above. A well-known M-theory corner in this duality web is DLCQ M-theory, where dimensionally reducing the lightlike circle gives M0T that contains BFSS. Via U-duality \cite{Obers:1998fb}, the lightlike circle in DLCQ M-theory maps to a spacelike circle in \emph{nonrelativistic M-theory} (NRMT) \cite{Gomis:2000bd} that uplifts NRST. 
NRMT is defined by the $\omega \rightarrow \infty$ limit obtained by the following parametrization of the background metric $G_\text{MN}$\,, $\text{M} = 0\,, \cdots, 10$ and gauge fields $A_3$ and $A_6$ \cite{Blair:2021waq, Ebert:2021mfu, Ebert:2023hba}:
\begin{subequations} \label{eq:nrmtrp}
\begin{align}
    & G^{}_\text{MN} = \omega^2 \, \tau^{}_\text{MN} + \omega^{-1} \, E^{}_\text{MN}\,, \\[4pt]
    & A^{}_3 = - \omega^3 \, \tau^0 \! \wedge \tau^1 \! \wedge \tau^2 \! + a^{}_3\,, 
        \,\, %
    A^{}_6 = {\tfrac{1}{2}} a^{}_3 \! \wedge \! A^{}_3  + a^{}_6\,, \label{eq:atas}
\end{align}
\end{subequations}
where $\tau^{}_\text{MN}$ and $E^{}_\text{MN}$ are in form the same as in Eq.~\eqref{eq:defte}. 
Moreover, the longitudinal sector is now 3D with $A = 0\,, 1\,, 2$ while the transverse sector is 8D with $A' = 3\,, \cdots, 10$\,. 
The U-duality which relates \eqref{eq:atas} to DLCQ M-theory acts on two longitudinal spatial directions and one transverse direction.
Dimensionally reducing NRMT over a shrinking two-torus
with one longitudinal cycle and one transverse cycle gives the S-dual IIB NRST and M1T~\cite{Ebert:2023hba}. Dimensionally reducing NRMT over a transverse circle gives M2T. 

U-duality acting on three longitudinal directions leads from \eqref{eq:nrmtrp} to the magnetic dual limit: 
\begin{subequations} \label{eq:nrmtrp6}
\begin{align}
    & G^{}_\text{MN} = \omega\, \tau^{}_\text{MN} + \omega^{-2} \, E^{}_\text{MN}\,, \\[4pt]
    & A^{}_3 =  a^{}_3\,, 
        \quad %
    A^{}_6 = - \omega^3 \, \tau^0 \! \wedge \dots \wedge \tau^5 \! + a^{}_6\,. 
\end{align}
\end{subequations}
For $\omega \rightarrow \infty$\,, this gives the critical $A_6$ limit of M-theory.
Via appropriate circle or toroidal oxidations, it provides the M-theory uplift of M$p$T for $p=3,4,5$. See \cite{Danielsson:2000gi, Gomis:2000bd, Gopakumar:2000ep,Bergshoeff:2000jn,Garcia:2002fa} for related decoupling limits. 

Finally, we consider the DLCQ of NRMT.
M(-1)T arises from dimensionally reducing DLCQ M-theory on a 2-torus with both cycles being lightlike, which connects to the M-theory uplifts of Carrollian strings. 
The DLCQ of NRMT uplifts MM0T 
and is U-dual to \emph{multicritical M-theory} (MMT) such that the lightlike circle is mapped to a spacelike circle along $\tilde{x}^{1}$. MMT uplifts MM1T and MM2T and is defined via the $\omega \rightarrow \infty$ limit of 
\begin{subequations}
\begin{align}
    G^{}_\text{MN} = & - \omega^2 \, \tau^{}_\text{M}{}^0 \, \tau^{}_\text{N}{}^0 + \Bigl( \tau^{}_\text{M}{}^1 \, \tau^{}_\text{N}{}^1 + \tau^{}_\text{M}{}^2 \, \tau^{}_\text{N}{}^2 \Bigr) \notag \\
    & + \omega \, \Bigl( \tau^{}_\text{M}{}^3 \, \tau^{}_\text{N}{}^3 + \tau^{}_\text{M}{}^{4} \, \tau^{}_\text{N}{}^{4} \Bigr) + \omega^{-1} \, E^{}_\text{MN}\,, \label{eq:mtusmt} \\[4pt]
    A_3 = & - \omega \, \tau^0 \wedge \tau^1 \wedge \tau^2 - \omega^2 \, \tau^0 \wedge \tau^3 \wedge \tau^4 + a_3\,,
\end{align}
\end{subequations}
with $A_6$ as in \eqref{eq:atas}. 
This multicritical limit arises from a critical background of two orthogonal M2-branes. 

\nicesection{Outlook}

The decoupling limits we have presented lead to corners of string and M-theory where the light excitations are (bound states of) branes, while other states decouple. 
We will study the derivation of these limits, the duality between them, and the physical description of these light excitations, in more detail in \cite{longpaper}.
We have also observed that the fundamental string worldsheet generically becomes nonrelativistic in these limits (except in NRST). The worldsheet in fact acquires the topology of nodal Riemann spheres, as in ambitwistor string theory~\cite{Geyer:2015bja}.
Worldsheet aspects of the duality web are detailed in \cite{Gomis:2023eav}.

Further extensions of the duality web are expected. 
Following the logic in this Letter, we could dualize MM$p$T and consider the limits from applying a third DLCQ and continuing to dualize.
Alternatively, the duality web of decoupling limits can be charted by using U-duality invariant BPS mass formulae~\cite{bpslimits}.
Restricting to M$p$T alone, there are  subtleties to deal with in the M$p$T limit for $p>3$, which in the Matrix theory literature required incorporating strong coupling behavior \cite{Rozali:1997cb, Berkooz:1997cq, Seiberg:1997ad}. For $p\geq 6$, this further involves confronting effects of low codimension branes, e.g.~M$6$T uplifts to a putative decoupling limit of M-theory associated with a background Kaluza-Klein monopole. Finally, the duality web of decoupling limits of half-maximal supersymmetric theories, e.g.~heterotic string theory \cite{Bergshoeff:2023fcf}, requires elucidation. 

The perspective of this Letter situates the Matrix Theory description of M-theory within a duality web of decoupling limits. 
It would be interesting to revisit aspects of Matrix theory, such as the precise correspondence with supergravity (see recent revival~\cite{Tropper:2023fjr,Herderschee:2023bnc,Herderschee:2023pza,Komatsu:2024vnb}), as well as Matrix descriptions beyond flat space in light of our better understanding of this picture. 
Crucially, the curved non-Lorentzian target space geometries should
be accompanied with constraints for consistency~\cite{longpaper}. This is motivated by study of NRST, where quantum consistency of the worldsheet~\cite{Yan:2021lbe} led to the imposition of the vanishing of components of the torsion $d\tau^A$. 
Such constraints were also required from a  
target space point of view in the $\mathcal{N}=1$ supersymmetric version of the limit~\cite{Bergshoeff:2021tfn}. 
It remains to be seen what constraints appear in the maximally supersymmetric case.

Matrix theory also plays a role in the AdS/CFT correspondence (for a recent reminder see \cite{Maldacena:2023acv}). For instance,  BFSS Matrix theory over a shrinking three-torus gives $\mathcal{N}=4$ super Yang-Mills in M3T, and is therefore expected to be associated with AdS${}_5$/CFT${}_4$. However, as we have argued above, M3T couples to a non-Lorentzian 3-brane Newton-Cartan spacetime geometry, which potentially points towards novel interplay between the boundary BPS behavior and the bulk AdS${}_5$ geometry. 

\vspace{3mm}

\noindent 
{\bf Acknowledgements:} We would like to thank E. Bergshoeff, J. de Boer, 
S. Ebert, T. Harmark, J. Gomis, K. Grosvenor, G. Oling, O. Schlotterer, B. Sundborg, K. Zarembo, R. Chatterjee for valuable discussions. The work of C.B. is supported through the Grants No. CEX2020-001007-S and No. PID2021-123017NB-I00, funded by MCIN/AEI/10.13039/501100011033 and by ERDF A way of making Europe. C.B. thanks Nordita for hospitality.  The work of N.O. is supported in part by VR project Grant 2021-04013 and Villum Foundation Experiment Project No. 00050317. 
Z.Y. is supported by the European Union’s Horizon 2020 research and innovation programme under the Marie Sk{\l}odowska-Curie Grant Agreement No. 31003710. Nordita is supported in part by NordForsk.

\bibliography{udlsmt}

\end{document}